# Incorporating action and reaction into a particle interpretation for quantum mechanics – Schrodinger case

Roderick I. Sutherland

Centre for Time, University of Sydney, NSW 2006 Australia

rod.sutherland@sydney.edu.au

**Abstract**

This paper follows on from a previous one in which it was shown that it is possible, within a de Broglie-Bohm style ontology for quantum mechanics, to incorporate action and reaction between the particle and its guiding field while remaining consistent with the usual experimental predictions. Whereas the previous paper focussed on the Dirac equation, the present work addresses the Schrodinger case and demonstrates that the same two-way interaction can be achieved. The aim in each case is to increase the physical plausibility of such models. The transition to include the reaction of the particle back on the field, and hence energy and momentum conservation, is attained by employing standard Lagrangian techniques. In formulating this description an interesting bonus emerges in that the hitherto unrelated concept of a gauge transformation is found to arise naturally as an essential part of the formalism. In particular, the phase S of the gauge transformation is seen to be the action function describing the hidden motion of the particle.

## 1. Introduction

This paper and a preceding one [1] relate to interpretations of quantum mechanics which take the underlying physical reality to consist of particles having definite and continuous trajectories, an obvious example of such an interpretation being the de Broglie-Bohm model [2,3]. The previous work dealt with the case of the Dirac equation and the present one extends the treatment to the Schrodinger equation. Although models proposing this picture of reality can reproduce the predictions of quantum mechanics, they contain an arguably artificial feature which is addressed here. In such models, agreement with experiment is achieved by assuming the particle is being influenced by a field, or pilot wave, which steers the particle's trajectory away from classical behaviour to the correct quantum predictions. This explanation of quantum phenomena, however, involves the questionable feature that the field acts on the particle but the particle does not act back on the field. A way to avoid this circumstance and incorporate action and reaction will be demonstrated here for the Schrodinger case.

In contrast to the usual way of formulating a particle interpretation for quantum mechanics, the novel ingredient here and in related papers [4-6] is the adoption of a Lagrangian approach. The advantages of such an approach are well known and include the fact that the field and particle equations of motion are both derivable from the same single expression and that, for a Lagrangian density with suitable symmetries, energy and momentum conservation is automatically ensured. Related to this last point is the fact that two-way interaction is automatically included as well. It therefore becomes a matter of finding an appropriate Lagrangian density expression which incorporates such an interaction while correctly reproducing the predictions of quantum mechanics. As shown previously in the Dirac paper and here again for the Schrodinger case, a suitable expression exists and can be employed. It describes the mutual influence via non-zero exchanges of energy and momentum between the particle and the field. In the present author's opinion, this two-way effect helps to make such an interpretation of quantum mechanics seem less contrived.

As a bonus, the formulation presented here provides a possible explanation for why the mathematics of quantum mechanics is "field-like" (i.e., involving partial differential equations) and yet experiments mainly detect "particle-like" events. It is found that the Lagrangian density considered here generates a field equation which contains source terms in addition to the usual field terms, indicating the presence of particles. This equation is, however, non-linear and would be unwieldy to employ. By incorporating the fact that a particle's position cannot be known precisely and continuously in quantum mechanics, a statistical form of the field equation is then formulated and it is found that the standard Schrodinger equation emerges. Although this familiar equation is linear and allows the usual predictions to be calculated more easily, the price paid is that the particle-like terms are thereby eliminated from the equation, making it appear as though physical reality consists only of fields.

The Schrodinger case considered here is straightforward but the mathematical presentation has proven to be somewhat longer than for the Dirac case. In particular, the field equation is seen to develop several more terms (which however, fortuitously cancel out). This extra complexity is mainly due to the fact that the Schrodinger case entails a second order differential equation whereas the Dirac equation is first order. As discussed in Sec. 2, previous work pursuing a Lagrangian approach to describe this sort of ontology has made it clear that only a very specific form of Lagrangian density expression can be compatible with the required quantum predictions. Fortunately this particular form continues to give favourable results.

The structure of the paper is as follows. Sections 2 to 4 summarise the basic Lagrangian formalism needed to accommodate action and reaction within non-relativistic, single-particle quantum mechanics. A general statistical framework is then introduced in Sec. 5 and a particular probability density is postulated in Sec. 6. Agreement with the Schrodinger equation is then demonstrated in Sec. 7. Since the logical steps are quite similar in the Dirac and Schrodinger cases, much of the wording is also similar in the two papers.

**2. Appropriate Lagrangian formalism**

As shown in the author's previous papers, an appropriate Lagrangian density leading to agreement with quantum mechanics can be obtained by working in analogy to the classical case of a charged particle interacting with an electromagnetic 4-potential $A^\alpha$. An essential step in pursuing the analogy is the replacement of the electromagnetic 4-vector $A^\alpha$ in the relevant Lagrangian formalism with another 4-vector which is a function of the quantum mechanical wavefunction $\psi$. To distinguish the new vector from $A^\alpha$, the notation $b^\alpha$ will be used here. The particular form chosen for the function $b^\alpha$ then depends on which wave equation is under consideration.

The resulting Lagrangian density $\mathcal{L}$ must describe both a particle and its guiding field. Some indication towards the appropriate expression for the Schrodinger case will be obtained here by briefly referring back to the relativistic case [1,4-6]. Previously, the necessary expression for $\mathcal{L}$ was found to conform for each wave equation to the following general form[1]:

$$\mathcal{L} = \mathcal{L}_{field} - \sigma_0 b (u_\alpha u^\alpha)^{1/2} + \sigma_0 u_\alpha b^\alpha \qquad (\alpha = 0,1,2,3) \qquad (1)$$

Here the term $\mathcal{L}_{field}$ is the usual textbook form of the Lagrangian density describing the wavefunction alone and the remaining two terms refer to the particle, with $u^\alpha$ being the particle's 4-velocity. The quantity $\sigma_0$ is the rest density distribution of the particle through space, with this distribution involving a delta function because the particle's "matter density" is concentrated at one point. The letter b is understood to represent the magnitude of the 4-vector $b^\alpha$:

$$b = (b_\alpha b^\alpha)^{1/2} \qquad (2)$$

For the Dirac case, the particular expression needed for $b^\alpha$ was found to be proportional to the usual 4-current density expression $\bar{\psi}\gamma^\alpha\psi$. The key feature which allows a model in agreement with quantum mechanics to be formulated is the fact that $\mathcal{L}$ involves both a vector $b^\alpha$ and its magnitude b in the way displayed in Eq. (1).

For the Schrodinger case to be considered here, a non-relativistic version of Eq. (1) is needed. The following analogous expression involving 3-vectors instead of 4-vectors will be found to yield the required results[2]:

$$\mathcal{L} = \mathcal{L}_{field} + \sigma b(-v_i v^i)^{1/2} + \sigma v_i b^i \qquad (i = 1,2,3) \qquad (3)$$

Here $v^i$ is the 3-velocity of the particle and $v_i = -v^i$. The "matter density" $\sigma$ of the particle is given by the delta function:

$$\sigma = \delta^3[\mathbf{x} - \mathbf{x}_p(t)] \qquad (4)$$

---

[1] In this equation, it is assumed that the metric tensor has signature $(+---)$ and there is summation over any repeated index.

[2] This expression differs from those in most classical mechanics texts in that it contains the factor $(-v_i v^i)^{1/2}$, rather than the usual form $-v_i v^i$. The ½ power is common in relativistic formulations, but not in the non-relativistic case. By starting with a general factor of the form $(-v_i v^i)^n$ or $(u_\alpha u^\alpha)^n$ in any of the Schrodinger, Dirac or Klein-Gordon cases, it can be shown that a model of the present sort will only work if the value $n = ½$ is chosen. It is interesting to note that this choice corresponds to "parameter invariance" in the relativistic case of Eq. (1), since the action function S for the particle can then be written as:

$$S = \int \left[ b(u_\alpha u^\alpha)^{1/2} + u_\alpha b^\alpha \right] d\lambda$$
$$= \int \left[ b\left(\frac{dx_\alpha}{d\lambda}\frac{dx^\alpha}{d\lambda}\right)^{1/2} + b^\alpha \frac{dx_\alpha}{d\lambda} \right] d\lambda$$
$$= \int \left[ b(dx_\alpha dx^\alpha)^{1/2} + b^\alpha dx_\alpha \right]$$

which is independent of the arbitrary parameter $\lambda$.

where $\mathbf{x}_p$ is the particle's spatial position as a function of time t and $\mathbf{x}$ is an arbitrary point in space. It is convenient at this stage to switch to the more familiar notation of non-relativistic physics (using boldface for vector quantities), with Eq. (3) then expressed as:

$$\mathcal{L} = \mathcal{L}_{field} + \sigma(bv - \mathbf{b} \cdot \mathbf{v}) \tag{5}$$

where:

$$b = |\mathbf{b}| = (\mathbf{b} \cdot \mathbf{b})^{1/2} \tag{6}$$

$$v = |\mathbf{v}| = (\mathbf{v} \cdot \mathbf{v})^{1/2} \tag{7}$$

The Lagrangian density expression in Eq. (5) is the one which will be employed through the rest of this paper. For the Schrodinger case, the particular form needed for the vector $\mathbf{b}$ in order to obtain agreement with quantum mechanics is found to be[3]:

$$\mathbf{b} = \frac{\hbar}{2i} \frac{\phi^* \vec{\nabla} \phi}{\phi^* \phi} \tag{8}$$

The letter $\phi$ has been used here rather than the usual $\psi$ because the field $\phi$ acting on the particle in this model is not directly equivalent to the wavefunction at this stage. The connection between these two functions will emerge shortly.

It will now be shown how the Lagrangian density in Eq. (5) leads to the usual wave equation.

**3. Field equation**

For the Schrodinger case of a particle with mass m, the field term in the overall Lagrangian density (5) is taken to be the following standard expression from text books[4]:

$$\mathcal{L}_{field} = -\frac{\hbar^2}{2m} \phi^* \vec{\nabla} \cdot \vec{\nabla} \phi + \frac{i\hbar}{2} \left( \phi^* \frac{\partial \phi}{\partial t} - \frac{\partial \phi^*}{\partial t} \phi \right) \tag{9}$$

A field equation for $\phi$ can now be obtained from Eq. (5) by performing a variation with respect to the complex conjugate field $\phi^*$. This procedure, although routine, is rather long and so has been relegated to an appendix. The following result is derived in Appendix 1:

$$-\frac{\hbar^2}{2m} \nabla^2 \phi - i\hbar \frac{\partial \phi}{\partial t} = \frac{i\hbar}{2} \nabla \cdot \left[ \frac{\sigma}{\phi^* \phi} \left( \mathbf{v} - \frac{v}{b} \mathbf{b} \right) \phi \right] + \frac{i\hbar}{2} \frac{\sigma}{\phi^* \phi} \left( \mathbf{v} - \frac{v}{b} \mathbf{b} \right) \cdot \nabla \phi + \frac{\sigma}{\phi^* \phi} \left( \mathbf{v} - \frac{v}{b} \mathbf{b} \right) \cdot \mathbf{b} \phi \tag{10}$$

The left hand side of this equation is seen to consist of the usual Schrodinger terms, whereas the right hand side can be viewed as consisting of source terms arising from the (hidden, but continuously existing) particle. It is, of course, necessary to explain why the right hand terms are not ruled out immediately by the existing experimental evidence. It is notable that this equation is more complicated than the Dirac case [1] because it contains three source terms rather than one.

---

[3] Here $\phi$ is a scalar field and $\phi^*$ is its complex conjugate, both being functions of position $\mathbf{x}$ and time t. The del operator $\vec{\nabla}$ is an abbreviation for $\vec{\nabla} - \overleftarrow{\nabla}$. Note that i is playing two roles in this paper, namely as the i[th] component of a vector and as the square root of minus one.

[4] See e.g., [7]. For simplicity, Eq. (9) has been limited to the free-space case and does not contain any external potentials. Terms describing an external scalar potential V and vector potential $\mathbf{A}$ can, however, easily be included in $\mathcal{L}_{field}$ without affecting the subsequent arguments.

In previous work [4-6], the field equations for the Dirac and Klein-Gordon cases were reduced to the standard wave equations by simply making the extra assumption that the round brackets such as those on the right hand side of Eq. (10) must be zero. This is the same as placing a restriction on the particle's velocity. In the present case, the round brackets would be zero if **v** were proportional to **b**. In particular, keeping Eq. (8) in mind, a natural choice would be:

$$\begin{aligned} \mathbf{v} &= \frac{1}{m}\mathbf{b} \\ &= \frac{\hbar}{2im}\frac{\phi^*\overrightarrow{\nabla}\phi}{\phi^*\phi} \end{aligned} \qquad (11)$$

This is identical to the guidance equation of the de Broglie-Bohm model once $\phi$ is assumed to be the usual wavefunction.

Eliminating the source terms in this way, however, also has the effect of removing any influence of the particle on the field, thereby ruling out any possibility of two-way interaction. Fortunately there is an alternative way of recovering the standard Schrodinger equation from the field equation (10) without this negative consequence. This other way of proceeding becomes available in the statistical case once it is acknowledged that the particle's position is not known precisely and needs to be described by a probability distribution. This approach is laid out from Sec. 5 onwards. As a preliminary step, however, the generalised momentum of the particle will now be derived so as to write Eq. (10) in a more convenient form.

**4. Generalised momentum**

The overall Lagrangian density in Eq. (5) can be expressed as the following sum:

$$\mathcal{L} = \mathcal{L}_{field} + \sigma L \qquad (12)$$

where L is the Lagrangian (as opposed to a Lagrangian density) governing the motion of the particle. Comparing Eqs. (5) and (12), the expression for L is:

$$L = bv - \mathbf{b}\cdot\mathbf{v} \qquad (13)$$

From this Lagrangian the particle's generalised momentum **p** can be derived via its standard definition, namely as the derivative of the Lagrangian with respect to velocity. As shown in Appendix 2, this leads to the following result[5]:

$$\mathbf{p} = \frac{b}{v}\mathbf{v} - \mathbf{b} \qquad (14)$$

This generalised momentum is also expressible via the following familiar Hamilton-Jacobi expression:

---

[5] It will be shown in Sec. 7 that the factor $b/v$ in this expression can be identified with the particle's mass m.

$$\mathbf{p} = \nabla S \tag{15}$$

where S is the action defined by[6]:

$$S(\mathbf{x}, t) = \int_{t_0}^{t} L \, dt \tag{16}$$

This last equation can be viewed as expressing the action integral for the particle's actual path as a function of the coordinates $(\mathbf{x}, t)$ at the upper limit of integration. Here L is again the Lagrangian given in Eq. (13) and $t_0$ is an arbitrary time in the past at which the particle's position $\mathbf{x}_0$ is taken to be fixed. Eqs. (14) and (15) can now be combined to give:

$$\nabla S = \frac{b}{v} \mathbf{v} - \mathbf{b} \tag{17}$$

and inserting this result back into the field equation (10) yields:

$$-\frac{\hbar^2}{2m} \nabla^2 \phi - i\hbar \frac{\partial \phi}{\partial t} = \frac{i\hbar}{2} \nabla \cdot \left[ \frac{\sigma}{\phi^* \phi} \left( \frac{v}{b} \nabla S \right) \phi \right] + \frac{i\hbar}{2} \frac{\sigma}{\phi^* \phi} \left( \frac{v}{b} \nabla S \right) \cdot \nabla \phi + \frac{\sigma}{\phi^* \phi} \left( \frac{v}{b} \nabla S \right) \cdot \mathbf{b} \, \phi \tag{18}$$

This form of the field equation will be more suitable for present requirements.

## 5. Statistical framework

It will now be highlighted by means of a well-known thought experiment that the usual wavefunction of quantum mechanics must be connected in only a statistical way to the field required in this model. Consider a point source emitting particles isotropically, so that each particle's wavefunction will evolve away from the source in a spherically symmetric fashion. If it is assumed that there is also a shell-like detector surrounding the point source at a certain radial distance, each particle will eventually be detected and will be seen to have travelled in its own particular direction, so that each particle breaks the spherical symmetry. Now the model proposed here entails that the particle is acting as a source of the field. It is therefore to be expected that the field, unlike the wavefunction, will be greater in the vicinity and direction of the path which the particle actually takes. This fact that the wavefunction will expand symmetrically but the field will not indicates that the two quantities can only be related statistically.

The statistical element which this argument requires will now be introduced by returning to the overall Lagrangian density (5) and taking a weighted average over the possible positions of the particle. The specific form of the position probability distribution is not immediately important and will be postponed until Sec. 6. Recalling from Eq. (4) that $\mathbf{x}_p$ is the particle's spatial position as a function of time t and that $\mathbf{x}$ is an arbitrary point in space, the desired weighted average can be obtained by multiplying the Lagrangian density (5) by the (as yet unknown) probability distribution $P(\mathbf{x}_p, t)$ and then integrating over $\mathbf{x}_p$. It will be assumed that the integral of $P(\mathbf{x}_p, t)$ over all space is equal to one. With expression (4) inserted for $\sigma$ and with the **x**'s and $\mathbf{x}_p$'s displayed explicitly, the following statistical version of the Lagrangian density is obtained:

$$\overline{\mathcal{L}}(\mathbf{x},t) = \iiint_{-\infty}^{+\infty} \left\{ \mathcal{L}_{\text{field}}(\mathbf{x},t) + \delta^3(\mathbf{x}-\mathbf{x}_p)[b(\mathbf{x},t)v - \mathbf{b}(\mathbf{x},t)\cdot\mathbf{v}] \right\} P(\mathbf{x}_p,t)\, d^3x_p \tag{19}$$

---

[6] Note that, unlike the standard de Broglie-Bohm model, S here is not equal to the phase of the wavefunction.

Performing the integrals then yields:

$$\overline{\mathcal{L}}(\mathbf{x},t) = \mathcal{L}_{\text{field}}(\mathbf{x},t) + P(\mathbf{x},t)[b(\mathbf{x},t)v - \mathbf{b}(\mathbf{x},t)\cdot\mathbf{v}] \tag{20}$$

Having obtained this more general expression, the next step is to find the field equation which it implies. Since $P(\mathbf{x},t)$ is independent of the field for the purposes of this derivation, the steps involved are essentially the same as those already carried out in Appendix 1, the only change being the replacement of $\sigma$ by $P(\mathbf{x},t)$. Therefore, by analogy with Eq. (18) above, the following result can be stated for the statistical version of the field equation:

$$-\frac{\hbar^2}{2m}\nabla^2\psi - i\hbar\frac{\partial\psi}{\partial t} = \frac{i\hbar}{2}\nabla\cdot\left[\frac{P(\mathbf{x},t)}{\psi^*\psi}\frac{v}{b}(\nabla S)\psi\right] + \frac{i\hbar}{2}\frac{P(\mathbf{x},t)}{\psi^*\psi}\frac{v}{b}(\nabla S)\cdot\nabla\psi + \frac{P(\mathbf{x},t)}{\psi^*\psi}\frac{v}{b}(\nabla S)\cdot\mathbf{b}\,\psi \tag{21}$$

The different letter $\psi$ has been used here rather than $\phi$ to highlight the fact that the field solution will now differ from that in previous sections because it is a solution of Eq. (21) instead of Eq. (18). The earlier solution is the actual field interacting with the particle whereas the new solution is the result obtained when only a probability distribution is inserted for the source particle's position, rather than a definite value[7].

## 6. Postulated probability distribution

In order to demonstrate consistency with standard quantum mechanics, a specific expression is now needed for the probability distribution $P(\mathbf{x},t)$. Such an expression cannot be derived and must be postulated separately, being more akin to a boundary condition than a basic part of the mathematical structure. It should, however, satisfy certain desirable conditions such as being conserved and being positive definite. An argument can be advanced in favour of a particular choice and, not surprisingly, this choice turns out to be the Born rule.

Conservation of probability at each point in space requires the distribution to satisfy a continuity equation and such a relationship can easily be derived from the field equation (21). Specifically, the desired form can be obtained by the familiar method of multiplying Eq. (21) on the left by $\psi^*$ and the complex conjugate of (21) on the right by $\psi$ then subtracting the resulting two equations. As shown in Appendix 3, this yields the following result:

$$\nabla\cdot\left\{P(\mathbf{x},t)\left[\frac{\hbar}{2im}\frac{\psi^*\overleftrightarrow{\nabla}\psi}{P(\mathbf{x},t)} + \frac{v}{b}\nabla S\right]\right\} + \frac{\partial(\psi^*\psi)}{\partial t} = 0 \tag{22}$$

---

[7] Note that the quantities b and $\nabla S$ in Eq. (21) remain defined by the expressions in Eqs. (6), (8) and (17) but with these expressions now written in terms of the latter solution $\psi$.

For comparison, the standard continuity equation for locally conserved probability has the form:

$$\nabla \cdot \{P(\mathbf{x},t)\mathbf{V}(\mathbf{x},t)\} + \frac{\partial P(\mathbf{x},t)}{\partial t} = 0 \tag{23}$$

where the upper case $\mathbf{V}$ here is the velocity of probability flow. In general, for an ensemble of particles described by a joint probability distribution for $\mathbf{x}$ and $\mathbf{v}$, the flow velocity $\mathbf{V}$ at each point $(\mathbf{x},t)$ is equal to the mean value $<\mathbf{v}>_{\mathbf{x},t}$ of the particle velocity $\mathbf{v}$ at that point.

Comparing Eqs. (22) and (23) it is seen that a viable choice for $P(\mathbf{x},t)$ would be:

$$P(\mathbf{x},t) = \psi^*\psi \tag{24}$$

The fact that $\psi^*\psi$ is positive definite lends further support to this choice. It will therefore be postulated that the particle's position should be described statistically by Eq. (24), which has the form of the familiar Born rule.

Further comparison of Eqs. (22) and (23) points to the following identification[8]:

$$\mathbf{V}(\mathbf{x},t) = \left[\frac{\hbar}{2im}\frac{\psi^*\vec{\nabla}\psi}{\psi^*\psi} + \frac{v}{b}\nabla S\right]$$

$$= \frac{1}{m}\mathbf{b} + \frac{v}{b}\nabla S \tag{25}$$

This expression will also be adopted and the fact that it can be usefully combined with an earlier equation for velocity will now be discussed.

The Hamilton-Jacobi formalism provides a specific solution for particle velocity $\mathbf{v}$ as a function of position $\mathbf{x}$ and time t. In particular, this formalism will provide a velocity solution for the Lagrangian density expression in Eq. (20). Such a solution entails there being only a single value of $\mathbf{v}$ at each point $(\mathbf{x},t)$ and hence the velocity $\mathbf{V}$ of probability flow reduces from an average to just this single value:

$$\mathbf{V}(\mathbf{x},t) = \mathbf{v}(\mathbf{x},t) \tag{26}$$

Now the relevant expression for $\mathbf{v}$ here[9] will satisfy the following rearranged version of Eq. (17):

$$\mathbf{v} = \frac{v}{b}\mathbf{b} + \frac{v}{b}\nabla S \tag{27}$$

Combining Eqs. (25), (26) and (27) then yields[10]:

---

[8] Note that $\mathbf{b}$ here is expressed in terms of the new solution $\psi$, as pointed out in footnote 7.

[9] Note that this $\mathbf{v}$ relates to the statistical case described by Eqs. (20) and (21) and so it is not the particle's actual velocity but merely the value calculated once there is only a statistical input. Nevertheless, this value is uniquely determined for each position and time, which justifies Eq. (26).

[10] Note that, although this relationship indicates that the **magnitude** of the $\mathbf{v}$ vector is proportional to that of the $\mathbf{b}$ vector, it does not imply that the vectors themselves are proportional. By writing (27) in the form

$$\frac{b}{v} = m \tag{28}$$

This result[11] and Eq. (24) will be used in the next section to simplify the field equation.

## 7. Reduction to the Schrodinger equation

By substituting Eqs. (24 and 28) into Eq. (21), the field equation is reduced to:

$$-\frac{\hbar^2}{2m}\nabla^2\psi - i\hbar\frac{\partial \psi}{\partial t} = \frac{i\hbar}{2}\nabla\cdot\left(\frac{\nabla S}{m}\psi\right) + \frac{i\hbar}{2}\frac{\nabla S}{m}\cdot\nabla\psi + \frac{1}{m}(\nabla S)\cdot\mathbf{b}\,\psi \tag{29}$$

In order to see the connection between this equation and the usual Schrodinger equation, two further identities which hold for the present Lagrangian formalism are needed at this point. These are derived in Appendix 4 and allow Eq. (29) to be adjusted to the following form:

$$-\frac{\hbar^2}{2m}\nabla^2\psi - i\hbar\frac{\partial \psi}{\partial t} = \frac{i\hbar}{2}\nabla\cdot\left(\frac{\nabla S}{m}\psi\right) + \frac{i\hbar}{2}\frac{\nabla S}{m}\cdot\nabla\psi - \frac{1}{2m}(\nabla S)^2\psi - \frac{\partial S}{\partial t}\psi \tag{30}$$

This version of the field equation still seems very different from the Schrodinger equation because of the non-zero terms on the right hand side – a total of four unwanted terms. However, a final step will clarify the situation. This step entails switching to a new field quantity via the following change of notation[12]:

$$\Psi(\mathbf{x},t) = \psi(\mathbf{x},t)\,e^{iS(\mathbf{x},t)/\hbar} \tag{31}$$

which is akin to performing a gauge transformation[13]. The quantity S in this equation is the same action as defined earlier and the upper case $\Psi$ will shortly be identified with the usual Schrodinger wavefunction. Note that S is real but $\psi$ and $\Psi$ will generally be complex. Now, as shown in Appendix 5, the change of notation (31) simplifies the field equation significantly. In particular, Eq. (30) is reduced to:

$$-\frac{\hbar^2}{2m}\nabla^2\Psi - i\hbar\frac{\partial \Psi}{\partial t} = 0 \tag{32}$$

which is just the standard Schrodinger equation. It has therefore been shown that the original Lagrangian density (5) leads, via a statistical treatment together with a familiar change of notation, to the correct wave equation in the Schrodinger case[14].

It is interesting to compare the relative advantages of the different field equations that have been derived here. The "non-statistical" Eq. (10) clearly contains both a field quantity $\phi$ and particle quantities $\sigma$ and $\mathbf{v}$ while giving a precise description of the influence (via the source

---

$\mathbf{v} = [\mathbf{b} + \nabla S(\mathbf{x},t)]/m$, it is seen that $\mathbf{v}$ and $\mathbf{b}$ generally point in different directions and that this difference varies with position and time.

[11] Eq. (28) also allows the generalised momentum in Eq. (14) to be written as $\mathbf{p} = m\mathbf{v} - \mathbf{b}$.

[12] Since there is no restriction imposed on S being larger than $\hbar$, the possibility arises that multiple S's could correspond to the same $\Psi$. This would, however, not affect the conclusions drawn here. In any case, the particle is acting as a source of the field in this model and so changing S is likely to cause a change in $\psi$ as well, so that $\Psi$ would not remain the same.

[13] The sign in the exponent here is opposite to the Dirac case.

[14] An analogous result can be derived for the Klein-Gordon case.

terms) of the particle on the field. This equation is non-linear, however, and would be difficult to solve. It would also need to be solved simultaneously with the particle's equation of motion. Turning to the "statistical" case of Eq. (32), which is just the standard Schrodinger equation, suddenly the equation is linear and can be solved relatively easily for $\Psi$ to obtain the usual quantum predictions. On the other hand, the price paid is that all evidence of a localised particle has been washed out of the equation, apparently indicating that fields are the only things which exist.

Up to this point the focus has been on the free-space case for simplicity. It is important, however, to highlight the further possibility which would arise if external potentials were included in the Schrodinger equation, specifically a scalar potential V and a vector potential **A**. In that more general case, the extra terms in Eq. (30) could disposed of by the alternative procedure of absorbing them into the potentials via a gauge-like transformation, i.e., by a change of notation for the potentials while keeping the lower case $\psi$ as the wavefunction.

The analysis in this paper has been concerned with showing how the particle can be still be influencing the field despite the apparent absence of source terms in the standard wave equation. For completeness, the continued presence of the effect in the other direction (i.e., field on particle) can be trivially confirmed by combining Eqs. (8), (27) and (28) to obtain the following velocity expression for the statistical case:

$$\mathbf{v} = \frac{1}{m}\left(\frac{\hbar}{2i}\frac{\psi^*\overleftrightarrow{\nabla}\psi}{\psi^*\psi} + \nabla S\right) \tag{33}$$

from which it is clear that the field is influencing the particle's velocity. Hence the influence is seen to be two-way[15].

Finally, as shown in Appendix 6, changing the notation of Eq. (33) via $\Psi = \psi e^{iS/\hbar}$ yields[16]:

$$\mathbf{v} = \frac{\hbar}{2im}\frac{\Psi^*\overleftrightarrow{\nabla}\Psi}{\Psi^*\Psi} \tag{34}$$

which is seen to be the same as the guidance equation for the de Broglie-Bohm model[17].

## 8. Discussion and Conclusions

Working within the context of a particle interpretation of quantum mechanics, a specific model has been constructed which incorporates action and reaction between the particle and the guiding field for the Schrodinger case. This model thereby demonstrates that two-way interaction can be achieved without contradicting the existing quantum predictions. Unlike in some of the author's preceding work [4-6], the back reaction of the particle on the field does not have the inconvenient property of reducing to zero in the special case of the quantum limit and continues unabated, although it becomes hidden from sight when the standard formalism of quantum mechanics is used.

---

[15] In the previous paper concerning the Dirac case, a detailed demonstration of energy and momentum conservation was given in terms of the zero 4-divergence of the system's energy-momentum tensor. This is also possible in the non-relativistic case but is somewhat more arduous and so will not be presented here.

[16] This change of notation also causes the probability distribution in Eq. (24) to be rewritten as $P(\mathbf{x},t) = \Psi^*\Psi$ (i.e., same expression but converted to upper case $\Psi$'s).

[17] Note that expression (34) is quite different from the expression tentatively mentioned in Eq. (11).

The model has resulted in two side benefits. First, it provides a possible explanation for why we seem to be dealing purely with propagating fields in the standard theory even though experiments generally detect particles. Second, it provides an unexpected physical link to the concept of a gauge transformation, which normally stands quite separately. On this latter point, the familiar gauge term $\nabla S$ is necessarily identified here as being the generalised momentum **p** of the associated particle.

Although the model gives rise to the standard Schrodinger equation and so to the usual predictions of quantum mechanics, there is some potential for it to make new predictions. This is because the extra feature introduced here of the particle influencing the field provides further scope for testable consequences to be devised.

**Appendix 1**

The wave equation for the field $\phi$ can be found most simply by applying the usual Lagrange formula [8], which here takes the form[18]:

$$\partial_\alpha \frac{\partial \mathcal{L}}{\partial(\partial_\alpha \phi^*)} - \frac{\partial \mathcal{L}}{\partial \phi^*} = 0 \qquad (\alpha = 0,1,2,3) \tag{35}$$

Now, from Eq. (12), the overall Lagrangian density $\mathcal{L}$ can be expressed as the following sum:

$$\mathcal{L} = \mathcal{L}_{\text{field}} + \sigma L \tag{36}$$

Substituting this result into Eq. (35) yields:

$$\left[\partial_\alpha \frac{\partial}{\partial(\partial_\alpha \bar{\phi})} - \frac{\partial}{\partial \bar{\phi}}\right] \mathcal{L}_{\text{field}} = -\left[\partial_\alpha \frac{\partial}{\partial(\partial_\alpha \bar{\phi})} - \frac{\partial}{\partial \bar{\phi}}\right] \sigma L \tag{37}$$

From Eq. (9), the field part of the Lagrangian density is:

$$\mathcal{L}_{\text{field}} = \frac{\hbar^2}{2m} \phi^* \overleftarrow{\partial}_j \overrightarrow{\partial}^j \phi + \frac{i\hbar}{2} \phi^* \overleftrightarrow{\partial}_0 \phi \qquad (j = 1,2,3) \tag{38}$$

and inserting this expression into the left hand side of Eq. (37) gives:

$$\left[\partial_\alpha \frac{\partial}{\partial(\partial_\alpha \phi^*)} - \frac{\partial}{\partial \phi^*}\right] \mathcal{L}_{\text{field}} = \left[\partial_i \frac{\partial}{\partial(\partial_i \phi^*)} + \partial_0 \frac{\partial}{\partial(\partial_0 \phi^*)} - \frac{\partial}{\partial \phi^*}\right] \left(\frac{\hbar^2}{2m} \phi^* \overleftarrow{\partial}_j \overrightarrow{\partial}^j \phi + \frac{i\hbar}{2} \phi^* \overleftrightarrow{\partial}_0 \phi\right)$$

$$= \partial_i \frac{\hbar^2}{2m} \delta^i_j \partial^j \phi - \frac{i\hbar}{2} \partial_0 \phi - \frac{i\hbar}{2} \partial_0 \phi$$

$$= \frac{\hbar^2}{2m} \partial_j \partial^j \phi - i\hbar \partial_0 \phi \tag{39}$$

---

[18] The symbol $\partial_\alpha$ here is an abbreviation for the partial derivative $\partial/\partial x^\alpha$. In particular, note that $\partial_0$ is being used from Eq. (39) onwards to represent $\partial/\partial t$.

To evaluate the right hand side of Eq. (37), the following expression for the Lagrangian L can be obtained from Eq. (13):

$$L = (-b_j b^j)^{1/2} v + v_j b^j \tag{40}$$

Noting that L's only dependence on $\phi$ and $\phi^*$ is via the vector $b^j$, the right hand side of (37) can be expressed in the form:

$$-\left[\partial_\alpha \frac{\partial}{\partial(\partial_\alpha \phi^*)} - \frac{\partial}{\partial \phi^*}\right]\sigma L = -\partial_\alpha\left[\sigma \frac{\partial L}{\partial b^j} \frac{\partial b^j}{\partial(\partial_\alpha \phi^*)}\right] + \sigma \frac{\partial L}{\partial b^j} \frac{\partial b^j}{\partial \phi^*} \tag{41}$$

An explicit expression for $b^j$ is now needed and this can be obtained by expressing Eq. (8) as follows:

$$b^j = \frac{i\hbar}{2} \frac{\phi^* \overleftrightarrow{\partial}^j \phi}{\phi^* \phi} \tag{42}$$

At this point it is convenient to introduce the following temporary abbreviations:

$$f^j = \frac{i\hbar}{2} \phi^* \overleftrightarrow{\partial}^j \phi \tag{43}$$

$$g = \phi^* \phi \tag{44}$$

so that $b^j$ can be expressed in the form:

$$b^j = \frac{f^j}{g} \tag{45}$$

This allows Eq (41) to be written as:

$$-\left[\partial_\alpha \frac{\partial}{\partial(\partial_\alpha \phi^*)} - \frac{\partial}{\partial \phi^*}\right]\sigma L$$

$$= -\partial_\alpha\left[\sigma \frac{\partial L}{\partial b^i}\left(\frac{\partial b^i}{\partial f^j}\frac{\partial f^j}{\partial(\partial_\alpha \phi^*)} + \frac{\partial b^i}{\partial g}\frac{\partial g}{\partial(\partial_\alpha \phi^*)}\right)\right] + \sigma \frac{\partial L}{\partial b^i}\left(\frac{\partial b^i}{\partial f^j}\frac{\partial f^j}{\partial \phi^*} + \frac{\partial b^i}{\partial g}\frac{\partial g}{\partial \phi^*}\right)$$

$$= -\partial_\alpha\left[\sigma \frac{\partial L}{\partial b^i}\left(\frac{1}{g}\frac{\partial f^i}{\partial f^j}\frac{\partial f^j}{\partial(\partial_\alpha \phi^*)} + f^i \frac{\partial(1/g)}{\partial g}\frac{\partial g}{\partial(\partial_\alpha \phi^*)}\right)\right] + \sigma \frac{\partial L}{\partial b^i}\left(\frac{1}{g}\frac{\partial f^i}{\partial f^j}\frac{\partial f^j}{\partial \phi^*} + f^i \frac{\partial(1/g)}{\partial g}\frac{\partial g}{\partial \phi^*}\right)$$

$$= -\partial_\alpha\left[\sigma \frac{\partial L}{\partial b^i}\left(\frac{1}{g}\delta^i_j\frac{\partial f^j}{\partial(\partial_\alpha \phi^*)} - \frac{f^i}{g^2}\frac{\partial g}{\partial(\partial_\alpha \phi^*)}\right)\right] + \sigma \frac{\partial L}{\partial b^i}\left(\frac{1}{g}\delta^i_j\frac{\partial f^j}{\partial \phi^*} - \frac{f^i}{g^2}\frac{\partial g}{\partial \phi^*}\right)$$

$$= -\partial_\alpha\left[\frac{\sigma}{g}\frac{\partial L}{\partial b^i}\left(\frac{\partial f^i}{\partial(\partial_\alpha \phi^*)} - b^i\frac{\partial g}{\partial(\partial_\alpha \phi^*)}\right)\right] + \frac{\sigma}{g}\frac{\partial L}{\partial b^i}\left(\frac{\partial f^i}{\partial \phi^*} - b^i\frac{\partial g}{\partial \phi^*}\right) \tag{46}$$

Inserting the expressions for $f^i$ and g from Eqs. (43) and (44), this then becomes:

$$-\left[\partial_\alpha \frac{\partial}{\partial(\partial_\alpha \phi^*)} - \frac{\partial}{\partial \phi^*}\right]\sigma L$$

$$= -\partial_\alpha\left[\frac{\sigma}{\phi^*\phi}\frac{\partial L}{\partial b^i}\left(\frac{\partial(\tfrac{1}{2}i\hbar\phi^*\overleftrightarrow{\partial}^i\phi)}{\partial(\partial_\alpha \phi^*)} - b^i \frac{\partial(\phi^*\phi)}{\partial(\partial_\alpha \phi^*)}\right)\right] + \frac{\sigma}{\phi^*\phi}\frac{\partial L}{\partial b^i}\left(\frac{\partial(\tfrac{1}{2}i\hbar\phi^*\overleftrightarrow{\partial}^i\phi)}{\partial \phi^*} - b^i\frac{\partial(\phi^*\phi)}{\partial \phi^*}\right)$$

$$= -\partial_\alpha\left[\frac{\sigma}{\phi^*\phi}\frac{\partial L}{\partial b^i}\left((-\tfrac{1}{2}i\hbar\, g^{\alpha i}\phi) - 0\right)\right] + \frac{\sigma}{\phi^*\phi}\frac{\partial L}{\partial b^i}\left(\tfrac{1}{2}i\hbar\,\partial^i\phi - b^i\phi\right)$$

$$= \frac{i\hbar}{2}\partial^i\left(\frac{\sigma}{\phi^*\phi}\frac{\partial L}{\partial b^i}\phi\right) + \frac{i\hbar}{2}\frac{\sigma}{\phi^*\phi}\frac{\partial L}{\partial b^i}\partial^i\phi - \frac{\sigma}{\phi^*\phi}\frac{\partial L}{\partial b^i}b^i\phi \qquad (47)$$

Now the derivative $\dfrac{\partial L}{\partial b^i}$ appearing in this equation can be evaluated via Eq. (40) as follows:

$$\frac{\partial L}{\partial b^i} = \frac{\partial}{\partial b^i}\left[(-b_j b^j)^{1/2}v + b_j v^j\right]$$

$$= \tfrac{1}{2}(-b_k b^k)^{-1/2}\left(-\frac{\partial b_j}{\partial b^i}b^j - b_j\frac{\partial b^j}{\partial b^i}\right)v + \frac{\partial b_j}{\partial b^i}v^j$$

$$= \tfrac{1}{2}(-b_k b^k)^{-1/2}(-g_{ij}b^j - b_j \delta^j_i)v + g_{ij}v^j$$

$$= \tfrac{1}{2}\frac{1}{b}(-2b_i)v + v_i$$

$$= v_i - \frac{v}{b}b_i \qquad (48)$$

Substituting this result into Eq. (47) then yields:

$$-\left[\partial_\alpha \frac{\partial}{\partial(\partial_\alpha\phi^*)} - \frac{\partial}{\partial\phi^*}\right]\sigma L = \frac{i\hbar}{2}\partial^i\left[\frac{\sigma}{\phi^*\phi}\left(v_i - \frac{v}{b}b_i\right)\phi\right] + \frac{i\hbar}{2}\frac{\sigma}{\phi^*\phi}\left(v_i - \frac{v}{b}b_i\right)\partial^i\phi - \frac{\sigma}{\phi^*\phi}\left(v_i - \frac{v}{b}b_i\right)b^i\phi$$
$$(49)$$

Finally, inserting the results (39) and (49) back into Eq. (37), the overall field equation for the Schrodinger case is found to be:

$$\frac{\hbar^2}{2m}\partial_i\partial^i\phi - i\hbar\partial_0\phi = \frac{i\hbar}{2}\partial_i\left[\frac{\sigma}{\phi^*\phi}\left(v^i - \frac{v}{b}b^i\right)\phi\right] + \frac{i\hbar}{2}\frac{\sigma}{\phi^*\phi}\left(v^i - \frac{v}{b}b^i\right)\partial_i\phi - \frac{\sigma}{\phi^*\phi}\left(v^i - \frac{v}{b}b^i\right)b_i\phi$$
$$(50)$$

or equivalently:

$$-\frac{\hbar^2}{2m}\nabla^2\phi - i\hbar\frac{\partial\phi}{\partial t} = \frac{i\hbar}{2}\nabla\cdot\left[\frac{\sigma_0}{\phi^*\phi}\left(\mathbf{v} - \frac{v}{b}\mathbf{b}\right)\phi\right] + \frac{i\hbar}{2}\frac{\sigma_0}{\phi^*\phi}\left(\mathbf{v} - \frac{v}{b}\mathbf{b}\right)\cdot\nabla\phi + \frac{\sigma_0}{\phi^*\phi}\left(\mathbf{v} - \frac{v}{b}\mathbf{b}\right)\cdot\mathbf{b}\,\phi$$
$$(51)$$

which completes the derivation of Eq. (10).

**Appendix 2**

From Eqs. (13) and (7), the Lagrangian L can be written in the form:

$$L = b(-v_j v^j)^{1/2} + b_j v^j \tag{52}$$

The particle's generalised momentum $p^i$ can then be derived from this expression as follows:

$$\begin{aligned}
p^i &= \frac{\partial L}{\partial v^i} \\
&= -\frac{\partial L}{\partial v_i} \\
&= -\frac{\partial}{\partial v_i}\left[ b(-v_j v^j)^{1/2} + b_j v^j \right] \\
&= -\left[ \tfrac{1}{2} b(-v_k v^k)^{-1/2} \left( -\frac{\partial v_j}{\partial v_i} v^j - v_j \frac{\partial v^j}{\partial v_i} \right) + b_j \frac{\partial v^j}{\partial v_i} \right] \\
&= \tfrac{1}{2} \frac{b}{v}\left( \delta^i_j v^j + v_j g^{ij} \right) - b_j g^{ij} \\
&= \frac{b}{v} v^i - b^i
\end{aligned} \tag{53}$$

i.e.: $\quad \mathbf{p} = \dfrac{b}{v}\mathbf{v} - \mathbf{b} \tag{54}$

which establishes Eq. (14).

**Appendix 3**

The complex conjugate of Eq. (21) is[19]:

$$-\frac{\hbar^2}{2m}\nabla^2 \psi^* + i\hbar \frac{\partial \psi^*}{\partial t} = -\frac{i\hbar}{2}\nabla\cdot\left[ \frac{P(\mathbf{x},t)}{\psi^*\psi}\frac{v}{b}(\nabla S)\psi^* \right] - \frac{i\hbar}{2}\frac{P(\mathbf{x},t)}{\psi^*\psi}\frac{v}{b}(\nabla S)\cdot\nabla\psi^* + \frac{P(\mathbf{x},t)}{\psi^*\psi}\frac{v}{b}(\nabla S)\cdot\mathbf{b}\,\psi^* \tag{55}$$

A continuity equation can be obtained from Eqs. (21) and (55) by multiplying (21) on the left by $\psi^*$ and (55) on the right by $\psi$ then subtracting the resulting two equations. This yields:

$$\begin{aligned}
-\frac{\hbar^2}{2m}\psi^*\nabla^2\psi &- i\hbar\psi^*\frac{\partial \psi}{\partial t} + \frac{\hbar^2}{2m}(\nabla^2\psi^*)\psi - i\hbar\frac{\partial\psi^*}{\partial t}\psi \\
&= \frac{i\hbar}{2}\psi^*\nabla\cdot\left[ \frac{P(\mathbf{x},t)}{\psi^*\psi}\frac{v}{b}(\nabla S)\psi \right] + \frac{i\hbar}{2}\frac{P(\mathbf{x},t)}{\psi^*\psi}\frac{v}{b}(\nabla S)\cdot\psi^*\nabla\psi \\
&\quad + \frac{i\hbar}{2}\nabla\cdot\left[ \frac{P(\mathbf{x},t)}{\psi^*\psi}\frac{v}{b}(\nabla S)\psi^* \right]\Psi + \frac{i\hbar}{2}\frac{P(\mathbf{x},t)}{\psi^*\psi}\frac{v}{b}(\nabla S)\cdot(\nabla\psi^*)\psi
\end{aligned} \tag{56}$$

---

[19] In taking the complex conjugate of Eq. (21), note that the quantities b and $\nabla S$ are real. This follows because **b** as defined in Eq. (8) is real and b and $\nabla S$ can then be expressed in terms of **b** via Eqs. (6) and (17), respectively.

The left hand side of this equation can be rewritten as follows:

$$-\frac{\hbar^2}{2m}\left[\nabla\cdot(\psi^*\nabla\psi)-(\nabla\psi^*)\cdot(\nabla\psi)\right]+\frac{\hbar^2}{2m}\left[\nabla\cdot(\psi\nabla\psi^*)-(\nabla\psi^*)\cdot(\nabla\psi)\right]-i\hbar\frac{\partial(\psi^*\psi)}{\partial t}$$
$$=-\frac{\hbar^2}{2m}\nabla\cdot(\psi^*\vec{\nabla}\psi)-i\hbar\frac{\partial(\psi^*\psi)}{\partial t} \qquad (57)$$

and the right hand side can be rewritten in the form:

$$\frac{i\hbar}{2}\psi^*\psi\nabla\cdot\left[\frac{P(\mathbf{x},t)}{\psi^*\psi}\frac{v}{b}\nabla S\right]+\frac{i\hbar}{2}\psi^*\frac{P(\mathbf{x},t)}{\psi^*\psi}\frac{v}{b}(\nabla S)\cdot\nabla\psi+\frac{i\hbar}{2}\frac{P(\mathbf{x},t)}{\psi^*\psi}\frac{v}{b}(\nabla S)\psi^*\cdot\nabla\psi$$
$$+\frac{i\hbar}{2}\psi^*\psi\nabla\cdot\left[\frac{P(\mathbf{x},t)}{\psi^*\psi}\frac{v}{b}\nabla S\right]+\frac{i\hbar}{2}\frac{P(\mathbf{x},t)}{\psi^*\psi}\frac{v}{b}(\nabla S)\cdot(\nabla\psi^*)\psi+\frac{i\hbar}{2}\frac{P(\mathbf{x},t)}{\psi^*\psi}\frac{v}{b}(\nabla S)\cdot(\nabla\psi^*)\psi$$
$$=i\hbar\psi^*\psi\nabla\cdot\left[\frac{P(\mathbf{x},t)}{\psi^*\psi}\frac{v}{b}\nabla S\right]+i\hbar\frac{P(\mathbf{x},t)}{\psi^*\psi}\frac{v}{b}(\nabla S)\cdot\nabla(\psi^*\psi)$$
$$=i\hbar\nabla\cdot\left[P(\mathbf{x},t)\frac{v}{b}\nabla S\right] \qquad (58)$$

Hence, combining Eqs. (57) and (58) yields:

$$-\frac{\hbar^2}{2m}\nabla\cdot(\psi^*\vec{\nabla}\psi)-i\hbar\frac{\partial(\psi^*\psi)}{\partial t}=i\hbar\nabla\cdot\left[P(\mathbf{x},t)\frac{v}{b}\nabla S\right] \qquad (59)$$

which rearranges to the form of a continuity equation:

$$\nabla\cdot\left\{P(\mathbf{x},t)\left[\frac{\hbar}{2im}\frac{\psi^*\vec{\nabla}\psi}{P(\mathbf{x},t)}+\frac{v}{b}\nabla S\right]\right\}+\frac{\partial(\psi^*\psi)}{\partial t}=0 \qquad (60)$$

This completes the derivation of Eq. (22).

**Appendix 4:**

Identity 1:

Eq. (17) can be rearranged to give:

$$v^i=\frac{v}{b}(b^i-\partial^i S) \qquad (61)$$

which then yields:

$$v_i v^i=\frac{v}{b}(b_i-\partial_i S)\frac{v}{b}(b^i-\partial^i S) \qquad (62)$$

i.e.:

$$-v^2=\frac{v^2}{b^2}\left[-b^2+(\partial_i S)(\partial^i S)-2(\partial_i S)b^i\right] \qquad (63)$$

This cancels to:

$$(\partial_i S)(\partial^i S) - 2(\partial_i S)b^i) = 0 \tag{64}$$

or equivalently:

$$(\nabla S) \cdot \mathbf{b} = -\tfrac{1}{2}(\nabla S) \cdot (\nabla S) \tag{65}$$

Identity 2:

From the definition of the action S in Eq. (16), one can write:

$$L = \frac{dS}{dt} \tag{66}$$

This expands to:

$$L = \frac{\partial S}{\partial x^i}\frac{\partial x^i}{\partial t} + \frac{\partial S}{\partial t}$$
$$= -p_i v^i + \frac{\partial S}{\partial t} \tag{67}$$

For the present Lagrangian in Eq. (13) this becomes:

$$-p_i v^i + \frac{\partial S}{\partial t} = bv + b_i v^i$$
$$= b\frac{(-v_i v^i)}{v} + b_i v^i$$
$$= -\left(\frac{b}{v}v_i - b_i\right)v^i$$
$$= -p_i v^i \qquad \text{(using Eq. (14))} \tag{68}$$

and so for this model the following condition holds:

$$\frac{\partial S}{\partial t} = 0 \tag{69}$$

This last result also implies the further simplification that the Hamiltonian is zero in this model.

Substituting the identities (65) and (69) into Eq. (29) then yields:

$$-\frac{\hbar^2}{2m}\nabla^2\psi - i\hbar\frac{\partial \psi}{\partial t} = \frac{i\hbar}{2}\nabla\cdot\left(\frac{\nabla S}{m}\psi\right) + \frac{i\hbar}{2}\frac{\nabla S}{m}\cdot\nabla\psi - \frac{1}{2m}(\nabla S)\cdot(\nabla S)\psi - \frac{\partial S}{\partial t}\psi \tag{70}$$

which is Eq. (30) as required.

**Appendix 5**

Using the change of notation $\Psi = \psi e^{iS/\hbar}$ introduced in Eq. (31), the following identity can be deduced:

$$\begin{aligned}
-\frac{\hbar^2}{2m}\nabla^2\Psi - i\hbar\frac{\partial \Psi}{\partial t} &= -\frac{\hbar^2}{2m}\nabla^2(\psi e^{iS/\hbar}) - i\hbar\frac{\partial}{\partial t}(\psi e^{iS/\hbar}) \\
&= -\frac{\hbar^2}{2m}\nabla\cdot\left[\left(\nabla\psi + \psi\frac{i}{\hbar}\nabla S\right)e^{iS/\hbar}\right] - i\hbar\left(\frac{\partial\psi}{\partial t} + \psi\frac{i}{\hbar}\frac{\partial S}{\partial t}\right)e^{iS/\hbar} \\
&= -\frac{\hbar^2}{2m}e^{iS/\hbar}\left[\nabla\cdot\left(\nabla + \frac{i}{\hbar}\nabla S\right)\psi + \frac{i}{\hbar}\nabla S\cdot\left(\nabla + \frac{i}{\hbar}\nabla S\right)\psi\right] - i\hbar e^{iS/\hbar}\left(\frac{\partial}{\partial t} + \frac{i}{\hbar}\frac{\partial S}{\partial t}\right)\psi \\
&= e^{iS/\hbar}\left\{-\frac{\hbar^2}{2m}\nabla^2\psi - i\hbar\frac{\partial\psi}{\partial t} - \frac{i\hbar}{2}\nabla\cdot\left(\frac{\nabla S}{m}\psi\right) - \frac{i\hbar}{2}\frac{\nabla S}{m}\cdot\nabla\psi + \frac{1}{2m}(\nabla S)^2\psi + \frac{\partial S}{\partial t}\psi\right\}
\end{aligned}$$
(71)

Now the field equation presented earlier in Eq. (30) can be written in the form:

$$-\frac{\hbar^2}{2m}\nabla^2\psi - i\hbar\frac{\partial\psi}{\partial t} - \frac{i\hbar}{2}\nabla\cdot\left(\frac{\nabla S}{m}\psi\right) - \frac{i\hbar}{2}\frac{\nabla S}{m}\cdot\nabla\psi + \frac{1}{2m}(\nabla S)^2\psi + \frac{\partial S}{\partial t}\psi = 0 \qquad (72)$$

This equation then implies that the curly bracket in Eq. (71) is zero. Therefore Eq. (71) reduces to:

$$-\frac{\hbar^2}{2m}\nabla^2\Psi - i\hbar\frac{\partial\Psi}{\partial t} = 0 \qquad (73)$$

which is the standard Schrodinger equation. This establishes the result stated in Eq. (32).

**Appendix 6**

Using $\Psi = \psi e^{iS/\hbar}$, Eq. (33) can be written equivalently in the form:

$$\begin{aligned}
\mathbf{v} &= \frac{\hbar}{2im}\frac{(\Psi e^{-iS/\hbar})^*\vec{\nabla}(\Psi e^{-iS/\hbar})}{(\Psi e^{-iS/\hbar})^*(\Psi e^{-iS/\hbar})} + \frac{\nabla S}{m} \\
&= \frac{\hbar}{2im\,\Psi^*\Psi}\left[\Psi^* e^{iS/\hbar}\left(\nabla\Psi - \Psi\frac{i}{\hbar}\nabla S\right)e^{-iS/\hbar} - \left(\nabla\Psi^* + \Psi^*\frac{i}{\hbar}\nabla S\right)e^{iS/\hbar}\Psi e^{-iS/\hbar}\right] + \frac{\nabla S}{m} \\
&= \frac{\hbar}{2im\,\Psi^*\Psi}\left[\Psi^*\vec{\nabla}\Psi - 2\Psi^*\Psi\frac{i}{\hbar}\nabla S\right] + \frac{\nabla S}{m} \\
&= \frac{\hbar}{2im}\frac{\Psi^*\vec{\nabla}\Psi}{\Psi^*\Psi}
\end{aligned}$$
(74)

as required for Eq. (34).

**References**

[1] R. I. Sutherland: *Incorporating action and reaction into a particle interpretation for quantum mechanics – Dirac case*. (2019). arXiv: 1908.04897.

[2] L. de Broglie: *Non-linear wave mechanics*. Elsevier, Amsterdam (1960).

[3] D. Bohm: *A suggested interpretation of quantum theory in terms of "hidden" variables*. I and II. Physical Review **85**, 166-179 and 180-193 (1952).

[4] R. I. Sutherland: *Lagrangian description for particle interpretations of quantum mechanics: single-particle case*. Foundations of Physics **45**, 1454-1464 (2015). arXiv:1411.3762.

[5] R. I. Sutherland: *Lagrangian description for particle interpretations of quantum mechanics: entangled many-particle case*. Foundations of Physics **47**, 174-207 (2017). arXiv:1509.02442.

[6] R. I. Sutherland: *A spacetime ontology compatible with quantum mechanics* Activitas Nervosa Superior **61**, 55-60 (2019). arXiv: 1904.05157.

[7] W. Greiner: *Relativistic Quantum Mechanics - Wave Equations*, 2nd edition, Springer Verlag, Berlin (1994).

[8] S. S. Schweber: *An Introduction to relativistic quantum field theory*, Harper International, New York (1964).